\documentclass[prl, twocolumn]{revtex4-1}
\usepackage{graphicx}
\usepackage{amssymb}
\usepackage[latin1]{inputenc}

\begin{document}
\title{Zero Landau level in folded graphene nanoribbons}
\author{E. Prada${}^1$, P. San-Jose${}^2$, L. Brey${}^1$}
\affiliation{${}^1$Instituto de Ciencia de Materiales de Madrid (CSIC), Cantoblanco, 28049 Madrid, Spain. \\${}^2$Instituto de Estructura de la Materia (CSIC), Serrano 123, 28006 Madrid, Spain}
\date{\today}
\begin{abstract}
Graphene nanoribbons can be folded into a double layer system keeping the two layers decoupled. In the Quantum Hall regime folds behave as a new type of Hall bar edge. We show that the symmetry properties of the zero Landau level in metallic nanoribbons dictate that the zero energy edge states traversing a fold are perfectly transmitted onto the opposite layer. This result is valid irrespective of fold geometry, magnetic field strength and crystallographic orientation of the nanoribbon. Backscattering suppression on the $N=0$ Hall plateau is ultimately due to the orthogonality of forward and backward channels, much like in the Klein paradox.
\end{abstract}

\maketitle

Graphene is a two-dimensional conductive membrane~\cite{Neto:RMP09} that can be folded into the third dimension without degradation to its electronic and structural properties. Folded graphene structures are recently becoming an experimental reality \cite{Huang:POTNAOS09}. Curled and folded edges might be much more ubiquitous than previously assumed \cite{Liu:PRL09, Fogler:10}. Their fascinating electrostatic \cite{Fogler:10,Feng:PRB09} and electronic properties \cite{Caetano:JCP08,Lammert:PRL00}, and the possibilities afforded by the third dimension, are rapidly making folded structures a very active graphene research subfield.  Folds under a uniform and perpendicular magnetic field are an ideal system to study effectively non-homogeneous magnetic fields, since the flux through the nanoribbon changes sign across the fold. This magnetic flux inversion can happen over very short lengthscales, something more difficult to achieve in semiconductor heterostructures.

Graphene monolayers under strong magnetic fields have the peculiarity of supporting a zero energy Landau level (LL) pinned at the Dirac point, where neutral graphene's particle and hole bands meet. The reason behind the pinning has been traced back to the Index Theorem and the Berry phase of chiral carriers moving under a magnetic field \cite{Nakahara:03,Novoselov:S04}. This apparently innocuous property has a host of profound consequences \cite{Prada:PRB07}. It is responsible for graphene's anomalous Quantum Hall effect (QHE) predicted \cite{Gusynin:PRL05} and observed \cite{Novoselov:N05} shortly after its discovery \cite{Novoselov:S04}.

\begin{figure}
\includegraphics[width=8cm]{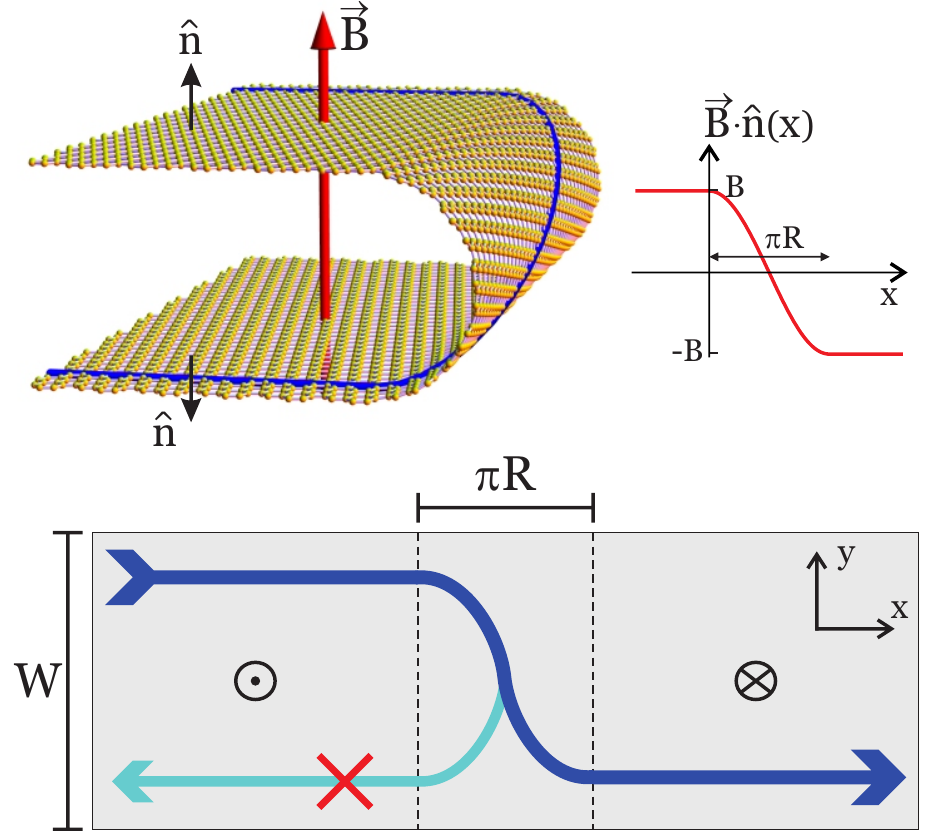}
\caption{\label{Fig:system}(Color online) Above: folded graphene nanoribbon in which the two layers are kept decoupled. A uniform magnetic field creates a magnetic flux that changes sign across the fold, like in a magnetic domain wall (below). The $N=0$ chiral edge state on the top layer is perfectly transmitted to the lower layer due to orthogonality of incoming and backscattered channels. 
}
\end{figure}

In this work we study the Quantum Hall physics in folded graphene nanoribbons, i.e., a double layer nanoribbon obtained by folding a monolayer so that the two layers remain decoupled (e.g. by an insulating buffer \cite{Schmidt:APL08}). We find that folds effectively become a new type of Hall bar edge along which the current flows in the Quantum Hall regime. Fold states, effectively chiral one-dimensional channels, are protected by topology just like edge states along terminated boundaries. In particular, we analyze the propagation of the $N=0$ edge state, or 'zero edge state' (ZES), across a fold. Such an edge state incoming towards the along the upper boundary of the top layer may either get 'transmitted' to the lower layer, or be 'reflected' back to the same layer, see Fig. \ref{Fig:system}. Both possibilities are in principle allowed by symmetry. Our main conclusion is that, due to orthogonality of incoming and backscattering channels, the ZES is however always perfectly transmitted to the opposite layer, irrespective of fold geometry, magnetic field, energy (within the zero Hall plateau) and nanoribbon crystallographic orientation, as long as the zero field nanoribbon is metallic. Only in the case of metallic armchair and at certain discrete energies close to the $N=1$ Landau level can resonant backscattering to the same layer take place. 

We consider the folded nanoribbon of width $W$ depicted in Fig. \ref{Fig:system}, with a fold of radius $R$. In the planar regions of the nanoribbon, the magnetic field $\vec{B}$ is either parallel or antiparallel to the nanoribbon's (oriented) normal. In the folded region the magnetic field has an in-plane component and a component normal to the ribbon, $\vec{B}\!\cdot\!\hat{n}$, that changes sign across the fold over a distance $\pi R$. Due to graphene's two-dimensional character, the orbital motion of the carriers in the ribbon, responsible for Landau levels and edge states, is only affected by the normal component. In the absence of Zeeman coupling, therefore, the fold is equivalent to a magnetic domain wall of width $\pi R$. Although the spin degeneracy of the carriers may be lifted by the Zeeman coupling to the total field, spin is conserved because $\vec B$ is constant in the embedding space, and trivially factors out of the scattering problem. 

In semiconductor systems in the Hall regime LLs carry a twofold spin ($s$) degeneracy (neglecting Zeeman).  The effect of boundaries on LLs is to make them dispersive. 
Graphene monolayers have an extra twofold `valley' ($\tau$) LL symmetry. Atomically sharp (`terminated') boundaries lift the LL valley degeneracy, forming split pairs of edge state branches in the dispersion relation (see Fig. S1 in the supplementary material). These branch pairs run roughly parallel to each other away from the Dirac point, so that at any given energy each LL contributes with two states per edge and spin. The zero LL, however, is different. It splits into two divergent branches of opposite energies, one particle and one hole-like, at either side of the Dirac point. As a consequence, it contributes with only half the number of edge states.

In two decoupled graphene layers, as those resulting from folding a monolayer nanoribbon onto itself without interlayer hybridization, LLs have an extra SU(2) layer degree of freedom ($\gamma$) that is also twofold degenerate, in addition to spin and valley. We can choose for example $\gamma=1$ for the upper layer in Fig. \ref{Fig:system}, and $\gamma=-1$ for the lower one. Although terminated edges break valley symmetry, they do not mix layers, which results in layer-degenerate edge states (along the horizontal boundaries in Fig. \ref{Fig:system}). Folds, in contrast, mix the two layers but do not break valley symmetry due to the lack of atomically sharp features. Therefore, layer degeneracy is lifted in fold states but valley degeneracy is preserved instead.

\begin{figure}
\includegraphics[width=8cm]{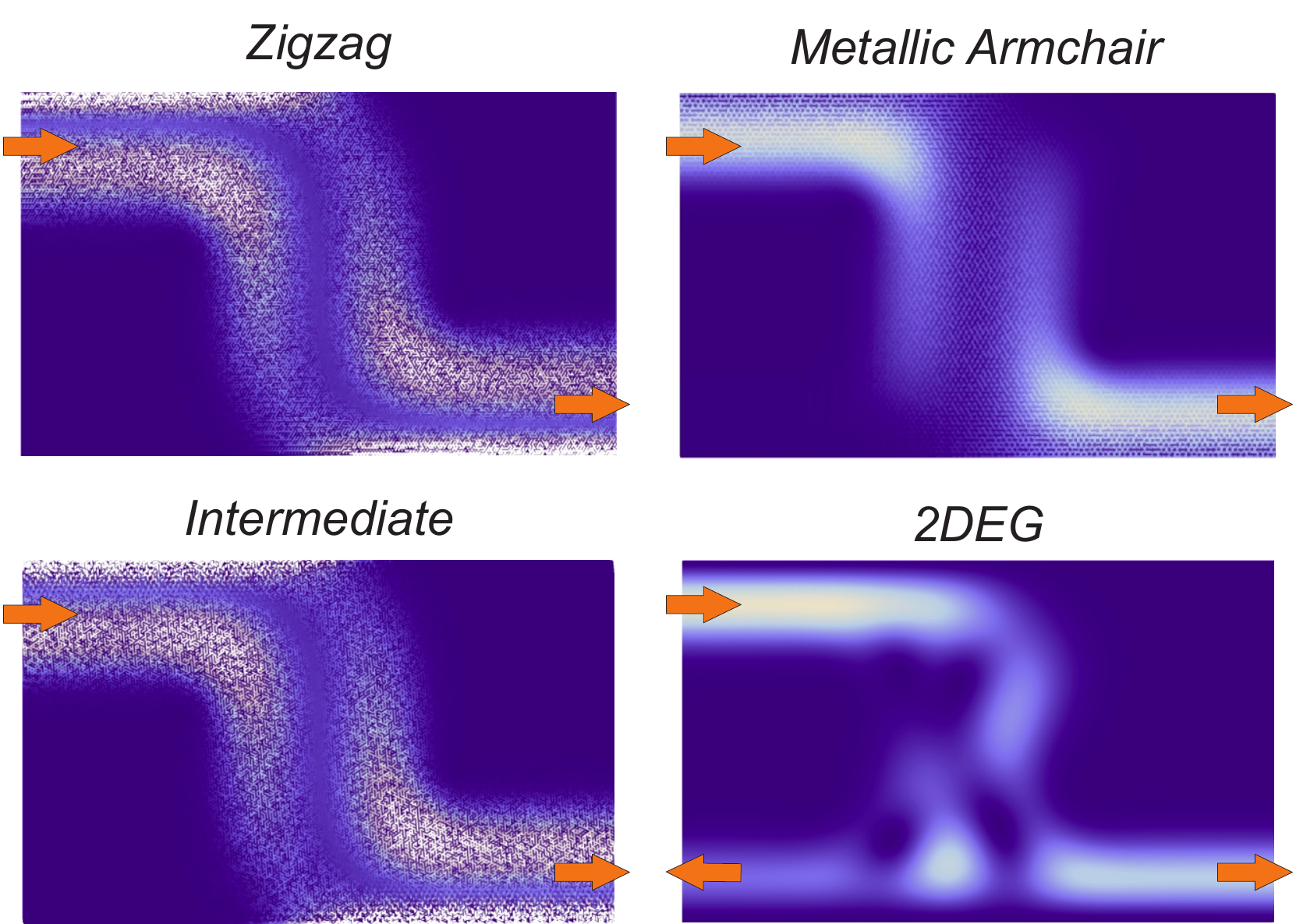}
\caption{\label{Fig:chargedensity} (Color online) Tight-binding numerical simulation of the charge density of the $N=0$ edge state as it propagates across the nanoribbon fold (central region of each plot) at energies close to the Dirac point. The nanoribbon is unfolded flat for the purpose of these plots. The case of a two dimensional electron gas is shown for comparison.}
\end{figure}

States in magnetic domain walls (and hence in folds) are carefully studied in Refs. \cite{Ghosh:PRB08,Park:PRB08,Oroszlany:PRB08}. They  are the quantum analogue of the classical snake states. In said works the magnetic flux profile is abrupt, which would correspond to our $R=0$ limit, and the flux interface is assumed to be infinitely long, $W\rightarrow \infty$. These assumptions allow to find analytical expressions for the energies and wavefunctions of the states formed around the magnetic domain wall. Some important features of these states are: (a) They are one dimensional channels confined between the gapped bulk of the left and right regions, with a bulk penetration length that depends on energy and magnetic length $l_B=(\hbar/|eB|)^{1/2}$. (b) They are chiral, i.e. they propagate only in one direction, just like edge states on terminated edges. (c) They populate both sides of the domain wall (i.e. both layers). In a folded nanoribbon $R$ and $W$ are finite and solutions require a numerical approach. However, the essential physics described above remains valid. Quasi-one-dimensional channels develop around the fold, but they smoothly connect to the terminated boundary states. Their penetration length is increased by approximately $\pi R$. In the central region they propagate from top to bottom in the choice of Fig. \ref{Fig:system}, and they populate both layers $\gamma=\pm 1$, see  vertical stripes on Fig. \ref{Fig:chargedensity}. These facts are independent of fold geometry (value of $R$ and precise shape of fold section), onsite potential variations or hopping amplitude modulations, for the same topological reasons as usual Hall edge states along terminated edges (change of the first Chern number respect to vacuum).

In the context of a transport problem in a fold structure, incoming edge states propagate along a given layer towards the fold. Due to the sign change of the magnetic flux, the electron is then forced to change direction and move along the fold and onto the opposite terminated edges. Nothing in principle precludes that the electron scatters onto an arbitrary superposition of the two layers after traversing the fold, see lower-right panel of Fig. \ref{Fig:chargedensity}. We show below, however, that the transmission of the ZES in particular is very simple and general: interlayer transmission is one, since backscattering onto the same layer becomes impossible by orthogonality of the initial and final states. We first provide an analytical proof for zigzag and armchair boundary terminations, and then present a numerical study for intermediate terminations which confirm this result in general.

It is well known \cite{Brey:PRB06a, Tworzydlo:PRB07, Akhmerov:PRB08a} that edge states along zigzag boundaries have a well defined (position independent) valley polarization, and belong to opposite valleys for opposite propagation directions as measured along the nanoribbon.
Denoting the upper/lower edges by $\alpha=\pm 1$, and the two layers by $\gamma=\pm 1$, the valley orientation $|\tau_{\alpha\gamma}\rangle$ of zigzag states is therefore 
 $|\tau_{\pm 1,\pm 1}\rangle=(1,0)$ and  $|\tau_{\mp 1,\pm 1}\rangle=(0,1)$. Incoming and outgoing states  on the same layer are valley-orthogonal. Likewise, it is straightforward to solve the low energy edge states with armchair boundary conditions \cite{Brey:PRB06a}. The resulting valley orientation is $|\tau_{\alpha\gamma}\rangle=(i \alpha \gamma e^{i\phi\gamma},1)/\sqrt{2}$. Here $e^{i\phi}=e^{i 2\pi (m+1)/3}$, where integer $m$ is the total number of sites across the ribbon. This number classifies the different armchair nanoribbons into metallic and semiconducting regarding their transport in the absence of a magnetic field. If $e^{i\phi}= 1$ (metallic armchair ribbons), the four valley orientations are pairwise orthogonal: $\langle \tau_{\alpha'\gamma'}|\tau_{\alpha\gamma}\rangle=\delta_{\alpha\gamma-\alpha'\gamma'}$, so opposite edges on the same layer are orthogonal, like in the zigzag case. Therefore, since valley is preserved along the fold propagation, backscattering onto the same layer is suppressed in both cases. In semiconducting armchair ribbons, in contrast, there is a finite overlap between all edge states, and backscattering can (and does) occur. Extending this analysis to intermediate crystallographic orientations is delicate. Valley polarization is no longer well defined, due to the contribution of a host of surface states, necessary to satisfy the atomic scale details of the boundary condition.

We have performed numerical simulations of the propagation of edge states on the hexagonal tight-binding nanoribbon of width $W$. The nanoribbon has parallel and minimal boundaries \cite{Akhmerov:PRB08a}, but arbitrary crystallographic orientation. A uniform magnetic field is applied to the folded nanoribbon structure by a Peierls substitution. We employ the recursive Green's function algorithm \cite{Datta:97}.  
 A double-sweep generalization \cite{Metalidis:PRB05} allows us to obtain the local charge density in real space, as seen in Fig. \ref{Fig:chargedensity}. Our numerical model remains valid as long as $R$ is large enough so as to neglect interlayer coupling, $\textrm{sp}_3$ orbital rehybridization and strains in the fold. This constraint is in practice not very stringent, and only requires that $R$ is much greater than the interatomic distance $a=1.4$\AA, the so called continuum limit. Folded structures as the one studied have no strain in the continuum limit, since the Gaussian curvature is zero, and therefore no pseudomagnetic field \cite{Neto:RMP09,Prada:PRB10} arises.
Regarding orbital rehybridization due to curvature, it has been shown to result in a coupling of mean curvature to the carrier densities, and the formation of electric dipoles \cite{Kim:EL08,Feng:PRB09}, which are not included in our model since they are only relevant for $R\sim a$. In any case, none of these perturbations can destroy fold states nor their chiral properties, which are protected by topology. 

In Fig. \ref{Fig:chargedensity} we plot some low energy results for the charge density distribution resulting from an edge state coming from the top layer (from the left in the domain wall picture) and propagating through the fold. We see that interlayer transfer of the zero energy state is perfect in the three ribbon terminations considered: armchair (parameterized by an angle $\theta=0$), zigzag ($\theta=\pi/6$) and an intermediate crystallographic orientation ($\theta=0.4 \pi/6$).  The parameters used for these simulations are: $W=81\textrm{nm}$, $R=16\textrm{nm}$, $l_B=8.1\textrm{nm}$, that corresponds to a magnetic field $B=10\textrm{T}$, and energy $\epsilon=10\textrm{meV}$. We have numerically %verified the results for arbitrary $W,R,l_B\gg a$.
checked that the results are valid for arbitrary $W,R,l_B\gg a$.

\begin{figure}
\includegraphics[width=8cm]{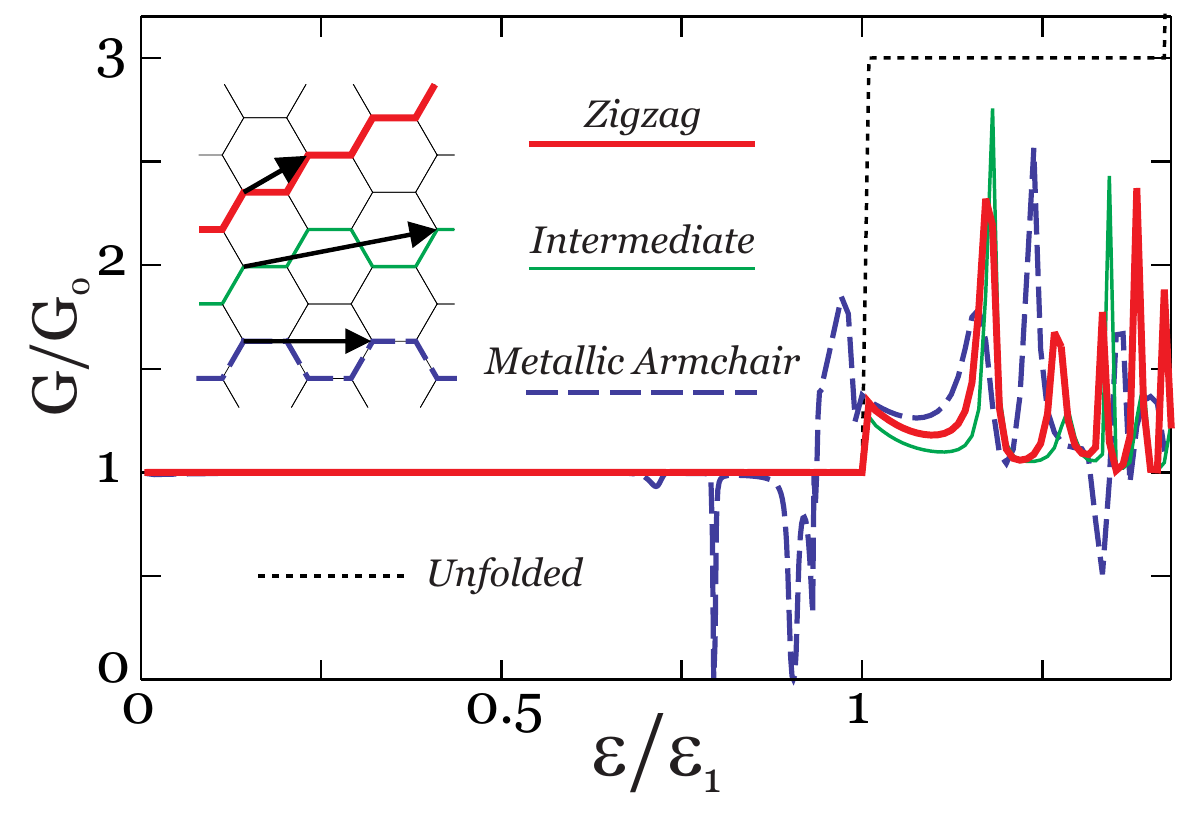}
\caption{\label{Fig:T} (Color online) Conductance versus Fermi energy $\epsilon$ across the nanoribbon fold for various boundary terminations. In the $N=0$ plateau, the $T=1$ observed for all orientations represents perfect layer-index transfer. Only at isolated energies in the metallic armchair case, quasi-bound states form and induce resonant backscattering.  The $N=1$ plateau has a transmission across the fold that is dominated by channel interference effects, which destroys the Quantum Hall plateaus.}
\end{figure}

In Fig. \ref{Fig:T} we present the ballistic two-terminal conductance $G=G_0\sum_\alpha T_\alpha$ (where $G_0=2e^2/h$ and $T_\alpha$ is the interlayer transmission of edge state $\alpha$). We plot $G$ as a function of the edge state energy, normalized to the first Landau level energy $\epsilon_1=\sqrt{2}\hbar v_F/\l_B$, for three different crystallographic orientations of the nanoribbon ($\theta=0$ for armchair, $\pi/6$ for zigzag and $0.1\pi/6$ for an intermediate orientation). We use the same $W$ and $l_B$ than in the previous figure, but here $R=20.3\textrm{nm}$. For $B=10\textrm{T}$, $\epsilon_1=0.1\textrm{eV}$. We have included for reference the curve for an unfolded zigzag nanoribbon (thin dotted line) of the same  width and with the same magnetic field. The ribbon with zigzag edges exhibits a perfect transmission $T=1$ throughout the $N=0$ plateau (thick red line). 
Other crystallographic orientations show identical behavior (thin green line), except for semiconducting armchair (not shown) that does exhibit backscattering. Metallic armchair ribbons (dashed blue line) may also exhibit backscattering at certain precise energies close to the $N=1$ subband (transmission dips). This occurs only in setups with $R$ greater than the magnetic length $l_B$, for which quasilocalized states can develop in the fold region. These are weakly hybridized with the ZES in the edges, and may produce sharp scattering resonances at which the pseudospin and valley quickly precess as energy crosses the quasi-bound state, see Fig. S2 in the supplementary material. This is known as resonant backscattering, and is a generic feature of quasi-1D quantum systems \cite{Gomez-Medina:PRL01}. Resonant backscattering requires, however, that the induced precession significantly modifies the ZES Dirac spinor structure in the fold, which only happens in the armchair case (see supplementary information). Backscattering on higher Hall plateaus is not suppressed. Since each edge supports more than one channel, they may mix along the fold, which leads to \emph{interchannel} backscattering. Indeed, at higher plateaus, transmission curves are quite complex, and are dominated by channel interference effects.

We have numerically observed that the transmission on the $N=0$ plateau is also one in the case of small magnetic fields, $l_B\lesssim W$, for which edge states are not well formed and span the whole ribbon width. The picture emerging from this result is that if an arbitrary clean nanoribbon without magnetic flux remains metallic as energy approaches zero, it will have perfect ZES interlayer transmission under a magnetic field when folded. Any such a ribbon has a level crossing of forward and backward channels at each Dirac point  that has not been lifted by its boundary conditions, and is also not lifted by adding an arbitrary space dependent magnetic field to the Dirac Hamiltonian. Since forward and backward scattering channels remain unmixed (unavoided crossing), backscattering is zero for the ZES. This result is independent of the details of the fold, so in particular it is also valid in other geometry variations, such as non-orthogonal folds, steps, double folds and loops (see Figs. S3 and S4 in the supplementary material).

Regarding disorder, only atomic scale defects close to the fold that can induce valley mixing of the ZES will affect the interlayer transmission. Consequently, it will tend to be suppressed by edge roughness. Our numerical simulations show that near-perfect transmission, however, persists at low energies for shallow edge roughness (see Fig. S5 in the supplementary material).

In conclusion, Hall regime transport through graphene nanoribbon folds implements, on the $N=0$ Hall plateau, a coherent $\pi$-rotation quantum gate on the layer index. In other words, a ZES propagating towards the fold from one layer is fully transferred to the opposite layer after crossing the fold. This is independent of energy, magnetic field strength, geometric details of the fold and crystallographic orientation of the nanoribbon (as long as the zero field nanoribbon is metallic). Isolated departures from this behavior only appear at certain resonant energies for the metallic nanoribbon with armchair edges. 
In the complementary picture of a magnetic domain wall, backscattering of ZESs is suppressed. This behavior is robust, and is a consequence of symmetry properties of the ZES wavefunctions, analogous to quasiparticle chirality in the absence of magnetic field. The phenomenon is reminiscent of the Klein paradox in graphene monolayers, in which backscattering of chiral quasiparticles on valley-preserving defects is suppressed because the normally reflected channel is orthogonal to the incoming one. Similarly, backscattering suppression of the ZES on a domain wall arises because the valley spinor of backward and forward channels are orthogonal, and the Dirac Hamiltonian in the fold does not couple valleys.

We wish to gratefully acknowledge the financial support from MICINN (Spain), through grants FIS2009-08744 (E.P and L.B) and FIS2008-00124 (P.S).

\bibliography{biblio}

%Merlin.mbs v4.21 2009-07-09.
\begin{thebibliography}{10}%
\makeatletter
\providecommand \@ifxundefined [1]{%
 \ifx #1\undefined \expandafter \@firstoftwo
 \else \expandafter \@secondoftwo
\fi
}%
\providecommand \@ifnum [1]{%
 \ifnum #1\expandafter \@firstoftwo
 \else \expandafter \@secondoftwo
\fi
}%
\providecommand \enquote [1]{``#1''}%
\providecommand \bibnamefont  [1]{#1}%
\providecommand \bibfnamefont [1]{#1}%
\providecommand \citenamefont [1]{#1}%
\providecommand\href[0]{\@sanitize\@href}%
\providecommand\@href[1]{\endgroup\@@startlink{#1}\endgroup\@@href}%
\providecommand\@@href[1]{#1\@@endlink}%
\providecommand \@sanitize [0]{\begingroup\catcode`\&12\catcode`\#12\relax}%
\@ifxundefined \pdfoutput {\@firstoftwo}{%
 \@ifnum{\z@=\pdfoutput}{\@firstoftwo}{\@secondoftwo}%
}{%
 \providecommand\@@startlink[1]{\leavevmode\special{html:<a href="#1">}}%
 \providecommand\@@endlink[0]{\special{html:</a>}}%
}{%
 \providecommand\@@startlink[1]{%
  \leavevmode
  \pdfstartlink
   attr{/Border[0 0 1 ]/H/I/C[0 1 1]}%
   user{/Subtype/Link/A<</Type/Action/S/URI/URI(#1)>>}%
  \relax
 }%
 \providecommand\@@endlink[0]{\pdfendlink}%
}%
\providecommand \url  [0]{\begingroup\@sanitize \@url }%
\providecommand \@url [1]{\endgroup\@href {#1}{\urlprefix}}%
\providecommand \urlprefix [0]{URL }%
\providecommand \Eprint[0]{\href }%
\@ifxundefined \urlstyle {%
  \providecommand \doi [1]{doi:\discretionary{}{}{}#1}%
}{%
  \providecommand \doi [0]{doi:\discretionary{}{}{}\begingroup
  \urlstyle{rm}\Url }%
}%
\providecommand \doibase [0]{http://dx.doi.org/}%
\providecommand \Doi[1]{\href{\doibase#1}}%
\providecommand \bibAnnote [3]{%
  \BibitemShut{#1}%
  \begin{quotation}\noindent
    \textsc{Key:}\ #2\\\textsc{Annotation:}\ #3%
  \end{quotation}%
}%
\providecommand \bibAnnoteFile [2]{%
  \IfFileExists{#2}{\bibAnnote {#1} {#2} {\input{#2}}}{}%
}%
\providecommand \typeout [0]{\immediate \write \m@ne }%
\providecommand \selectlanguage [0]{\@gobble}%
\providecommand \bibinfo [0]{\@secondoftwo}%
\providecommand \bibfield [0]{\@secondoftwo}%
\providecommand \translation [1]{[#1]}%
\providecommand \BibitemOpen[0]{}%
\providecommand \bibitemStop [0]{}%
\providecommand \bibitemNoStop [0]{.\EOS\space}%
\providecommand \EOS [0]{\spacefactor3000\relax}%
\providecommand \BibitemShut [1]{\csname bibitem#1\endcsname}%
%</preamble>
\bibitem{Neto:RMP09}%
  \BibitemOpen
  \bibfield{author}{%
  \bibinfo {author} {\bibfnamefont{A.~H.~C.}\ \bibnamefont{Neto}}, \bibinfo
  {author} {\bibfnamefont{F.}~\bibnamefont{Guinea}}, \bibinfo {author}
  {\bibfnamefont{N.~M.~R.}\ \bibnamefont{Peres}}, \bibinfo {author}
  {\bibfnamefont{K.~S.}\ \bibnamefont{Novoselov}},\ and\ \bibinfo {author}
  {\bibfnamefont{A.~K.}\ \bibnamefont{Geim}},\ }%
  \bibfield{journal}{%
  \bibinfo {journal} {Rev. Mod. Phys.}\ }%
  \textbf{\bibinfo {volume} {81}},\ \bibinfo {pages} {109} (\bibinfo {year}
  {2009})%
  \bibAnnoteFile{NoStop}{Neto:RMP09}%
\bibitem{Huang:POTNAOS09}%
  \BibitemOpen
  \bibfield{author}{%
  \bibinfo {author} {\bibfnamefont{J.}~\bibnamefont{Huang}}, \bibinfo {author}
  {\bibfnamefont{F.}~\bibnamefont{Ding}}, \bibinfo {author}
  {\bibfnamefont{B.}~\bibnamefont{Yakobson}}, \bibinfo {author}
  {\bibfnamefont{P.}~\bibnamefont{Lu}}, \bibinfo {author}
  {\bibfnamefont{L.}~\bibnamefont{Qi}},\ and\ \bibinfo {author}
  {\bibfnamefont{J.}~\bibnamefont{Li}},\ }%
  \bibfield{journal}{%
  \bibinfo {journal} {Proc. Nat. Ac. Sci.}\ }%
  \textbf{\bibinfo {volume} {106}},\ \bibinfo {pages} {10103} (\bibinfo {year}
  {2009})%
  \bibAnnoteFile{NoStop}{Huang:POTNAOS09}%
\bibitem{Liu:PRL09}%
  \BibitemOpen
  \bibfield{author}{%
  \bibinfo {author} {\bibfnamefont{Z.}~\bibnamefont{Liu}}, \bibinfo {author}
  {\bibfnamefont{K.}~\bibnamefont{Suenaga}}, \bibinfo {author}
  {\bibfnamefont{P.~J.~F.}\ \bibnamefont{Harris}},\ and\ \bibinfo {author}
  {\bibfnamefont{S.}~\bibnamefont{Iijima}},\ }%
  \bibfield{journal}{%
  \bibinfo {journal} {Phys. Rev. Lett.}\ }%
  \textbf{\bibinfo {volume} {102}},\ \bibinfo {pages} {015501} (\bibinfo {year}
  {2009})%
  \bibAnnoteFile{NoStop}{Liu:PRL09}%
\bibitem{Fogler:10}%
  \BibitemOpen
  \bibfield{author}{%
  \bibinfo {author} {\bibfnamefont{M.}~\bibnamefont{Fogler}}, \bibinfo {author}
  {\bibfnamefont{A.}~\bibnamefont{Neto}},\ and\ \bibinfo {author}
  {\bibfnamefont{F.}~\bibnamefont{Guinea}},\ }%
  \bibinfo {journal} {arXiv:1002.3418}%
  \bibAnnoteFile{NoStop}{Fogler:10}%
\bibitem{Feng:PRB09}%
  \BibitemOpen
\bibfield{journal}{%
    }%
  \bibfield{author}{%
  \bibinfo {author} {\bibfnamefont{J.}~\bibnamefont{Feng}}, \bibinfo {author}
  {\bibfnamefont{L.}~\bibnamefont{Qi}}, \bibinfo {author}
  {\bibfnamefont{J.~Y.}\ \bibnamefont{Huang}},\ and\ \bibinfo {author}
  {\bibfnamefont{J.}~\bibnamefont{Li}},\ }%
  \bibfield{journal}{%
  \bibinfo {journal} {Phys. Rev. B}\ }%
  \textbf{\bibinfo {volume} {80}},\ \bibinfo {pages} {165407} (\bibinfo {year}
  {2009})%
  \bibAnnoteFile{NoStop}{Feng:PRB09}%
\bibitem{Caetano:JCP08}%
  \BibitemOpen
  \bibfield{author}{%
  \bibinfo {author} {\bibfnamefont{E.~W.~S.}\ \bibnamefont{Caetano}}, \bibinfo
  {author} {\bibfnamefont{V.~N.}\ \bibnamefont{Freire}}, \bibinfo {author}
  {\bibfnamefont{S.~G.}\ \bibnamefont{dos Santos}}, \bibinfo {author}
  {\bibfnamefont{D.~S.}\ \bibnamefont{Galv{\~a}o}},\ and\ \bibinfo {author}
  {\bibfnamefont{F.}~\bibnamefont{Satoa}},\ }%
  \bibfield{journal}{%
  \bibinfo {journal} {J. Chem. Phys.}\ }%
  \textbf{\bibinfo {volume} {128}},\ \bibinfo {pages} {164719} (\bibinfo {year}
  {2008})%
  \bibAnnoteFile{NoStop}{Caetano:JCP08}%
\bibitem{Lammert:PRL00}%
  \BibitemOpen
  \bibfield{author}{%
  \bibinfo {author} {\bibfnamefont{P.~E.}\ \bibnamefont{Lammert}}, \bibinfo
  {author} {\bibfnamefont{P.}~\bibnamefont{Zhang}},\ and\ \bibinfo {author}
  {\bibfnamefont{V.~H.}\ \bibnamefont{Crespi}},\ }%
  \bibfield{journal}{%
  \bibinfo {journal} {Physical Review Letters}\ }%
  \textbf{\bibinfo {volume} {84}},\ \bibinfo {pages} {2453} (\bibinfo {year}
  {2000})%
  \bibAnnoteFile{NoStop}{Lammert:PRL00}%
\bibitem{Nakahara:03}%
  \BibitemOpen
  \bibfield{author}{%
  \bibinfo {author} {\bibfnamefont{M.}~\bibnamefont{Nakahara}},\ }%
  \emph{\bibinfo {title} {Geometry, topology, and physics}}\ (\bibinfo
  {publisher} {Taylor \& Francis},\ \bibinfo {year} {2003})%
  \bibAnnoteFile{NoStop}{Nakahara:03}%
\bibitem{Novoselov:S04}%
  \BibitemOpen
  \bibfield{author}{%
  \bibinfo {author} {\bibfnamefont{K.}~\bibnamefont{Novoselov}}, \bibinfo
  {author} {\bibfnamefont{A.}~\bibnamefont{Geim}}, \bibinfo {author}
  {\bibfnamefont{S.}~\bibnamefont{Morozov}}, \bibinfo {author}
  {\bibfnamefont{D.}~\bibnamefont{Jiang}}, \bibinfo {author}
  {\bibfnamefont{Y.}~\bibnamefont{Zhang}}, \bibinfo {author}
  {\bibfnamefont{S.}~\bibnamefont{Dubonos}}, \bibinfo {author}
  {\bibfnamefont{I.}~\bibnamefont{Grigorieva}},\ and\ \bibinfo {author}
  {\bibfnamefont{A.}~\bibnamefont{Firsov}},\ }%
  \bibfield{journal}{%
  \bibinfo {journal} {Science}\ }%
  \textbf{\bibinfo {volume} {306}},\ \bibinfo {pages} {666} (\bibinfo {year}
  {2004})%
  \bibAnnoteFile{NoStop}{Novoselov:S04}%
\bibitem{Prada:PRB07}%
  \BibitemOpen
  \bibfield{author}{%
  \bibinfo {author} {\bibfnamefont{E.}~\bibnamefont{Prada}}, \bibinfo {author}
  {\bibfnamefont{P.}~\bibnamefont{San-Jose}}, \bibinfo {author}
  {\bibfnamefont{B.}~\bibnamefont{Wunsch}},\ and\ \bibinfo {author}
  {\bibfnamefont{F.}~\bibnamefont{Guinea}},\ }%
  \bibfield{journal}{%
  \bibinfo {journal} {Phys. Rev. B}\ }%
  \textbf{\bibinfo {volume} {75}},\ \bibinfo {pages} {113407} (\bibinfo {year}
  {2007})%
  \bibAnnoteFile{NoStop}{Prada:PRB07}%
\bibitem{Gusynin:PRL05}%
  \BibitemOpen
  \bibfield{author}{%
  \bibinfo {author} {\bibfnamefont{V.~P.}\ \bibnamefont{Gusynin}}\ and\
  \bibinfo {author} {\bibfnamefont{S.~G.}\ \bibnamefont{Sharapov}},\ }%
  \bibfield{journal}{%
  \bibinfo {journal} {Phys. Rev. Lett.}\ }%
  \textbf{\bibinfo {volume} {95}},\ \bibinfo {pages} {146801} (\bibinfo {year}
  {2005})%
  \bibAnnoteFile{NoStop}{Gusynin:PRL05}%
\bibitem{Novoselov:N05}%
  \BibitemOpen
  \bibfield{author}{%
  \bibinfo {author} {\bibfnamefont{K.}~\bibnamefont{Novoselov}}, \bibinfo
  {author} {\bibfnamefont{A.}~\bibnamefont{Geim}}, \bibinfo {author}
  {\bibfnamefont{S.}~\bibnamefont{Morozov}}, \bibinfo {author}
  {\bibfnamefont{D.}~\bibnamefont{Jiang}}, \bibinfo {author}
  {\bibfnamefont{M.}~\bibnamefont{Katsnelson}}, \bibinfo {author}
  {\bibfnamefont{I.}~\bibnamefont{Grigorieva}}, \bibinfo {author}
  {\bibfnamefont{S.}~\bibnamefont{Dubonos}},\ and\ \bibinfo {author}
  {\bibfnamefont{A.}~\bibnamefont{Firsov}},\ }%
  \bibfield{journal}{%
  \bibinfo {journal} {Nature}\ }%
  \textbf{\bibinfo {volume} {438}},\ \bibinfo {pages} {197} (\bibinfo {year}
  {2005})%
  \bibAnnoteFile{NoStop}{Novoselov:N05}%
\bibitem{Schmidt:APL08}%
  \BibitemOpen
  \bibfield{author}{%
  \bibinfo {author} {\bibfnamefont{H.}~\bibnamefont{Schmidt}}, \bibinfo
  {author} {\bibfnamefont{T.}~\bibnamefont{L{\"u}dtke}}, \bibinfo {author}
  {\bibfnamefont{P.}~\bibnamefont{Barthold}}, \bibinfo {author}
  {\bibfnamefont{E.}~\bibnamefont{McCann}}, \bibinfo {author}
  {\bibfnamefont{V.}~\bibnamefont{Fal'ko}},\ and\ \bibinfo {author}
  {\bibfnamefont{R.}~\bibnamefont{Haug}},\ }%
  \bibfield{journal}{%
  \bibinfo {journal} {Appl. Phys. Lett.}\ }%
  \textbf{\bibinfo {volume} {93}},\ \bibinfo {pages} {172108} (\bibinfo {year}
  {2008})%
  \bibAnnoteFile{NoStop}{Schmidt:APL08}%
\bibitem{Ghosh:PRB08}%
  \BibitemOpen
  \bibfield{author}{%
  \bibinfo {author} {\bibfnamefont{T.~K.}\ \bibnamefont{Ghosh}}, \bibinfo
  {author} {\bibfnamefont{A.}~\bibnamefont{De~Martino}}, \bibinfo {author}
  {\bibfnamefont{W.}~\bibnamefont{H{\"a}usler}}, \bibinfo {author}
  {\bibfnamefont{L.}~\bibnamefont{Dell'Anna}},\ and\ \bibinfo {author}
  {\bibfnamefont{R.}~\bibnamefont{Egger}},\ }%
  \bibfield{journal}{%
  \bibinfo {journal} {Phys. Rev. B}\ }%
  \textbf{\bibinfo {volume} {77}},\ \bibinfo {pages} {081404(R)} (\bibinfo
  {year} {2008})%
  \bibAnnoteFile{NoStop}{Ghosh:PRB08}%
\bibitem{Park:PRB08}%
  \BibitemOpen
  \bibfield{author}{%
  \bibinfo {author} {\bibfnamefont{S.}~\bibnamefont{Park}}\ and\ \bibinfo
  {author} {\bibfnamefont{H.~S.}\ \bibnamefont{Sim}},\ }%
  \bibfield{journal}{%
  \bibinfo {journal} {Phys. Rev. B}\ }%
  \textbf{\bibinfo {volume} {77}},\ \bibinfo {pages} {075433} (\bibinfo {year}
  {2008})%
  \bibAnnoteFile{NoStop}{Park:PRB08}%
\bibitem{Oroszlany:PRB08}%
  \BibitemOpen
  \bibfield{author}{%
  \bibinfo {author} {\bibfnamefont{L.}~\bibnamefont{Oroszl\'any}}, \bibinfo
  {author} {\bibfnamefont{P.}~\bibnamefont{Rakyta}}, \bibinfo {author}
  {\bibfnamefont{A.}~\bibnamefont{Korm\'anyos}}, \bibinfo {author}
  {\bibfnamefont{C.~J.}\ \bibnamefont{Lambert}},\ and\ \bibinfo {author}
  {\bibfnamefont{J.}~\bibnamefont{Cserti}},\ }%
  \bibfield{journal}{%
  \bibinfo {journal} {Phys. Rev. B}\ }%
  \textbf{\bibinfo {volume} {77}},\ \bibinfo {pages} {081403(R)} (\bibinfo
  {year} {2008})%
  \bibAnnoteFile{NoStop}{Oroszlany:PRB08}%
\bibitem{Brey:PRB06a}%
  \BibitemOpen
  \bibfield{author}{%
  \bibinfo {author} {\bibfnamefont{L.}~\bibnamefont{Brey}}\ and\ \bibinfo
  {author} {\bibfnamefont{H.~A.}\ \bibnamefont{Fertig}},\ }%
  \bibfield{journal}{%
  \bibinfo {journal} {Phys. Rev. B}\ }%
  \textbf{\bibinfo {volume} {73}},\ \bibinfo {pages} {195408} (\bibinfo {year}
  {2006})%
  \bibAnnoteFile{NoStop}{Brey:PRB06a}%
\bibitem{Tworzydlo:PRB07}%
  \BibitemOpen
  \bibfield{author}{%
  \bibinfo {author} {\bibfnamefont{J.}~\bibnamefont{Tworzydlo}}, \bibinfo
  {author} {\bibfnamefont{I.}~\bibnamefont{Snyman}}, \bibinfo {author}
  {\bibfnamefont{A.~R.}\ \bibnamefont{Akhmerov}},\ and\ \bibinfo {author}
  {\bibfnamefont{C.~W.~J.}\ \bibnamefont{Beenakker}},\ }%
  \bibfield{journal}{%
  \bibinfo {journal} {Phys. Rev. B}\ }%
  \textbf{\bibinfo {volume} {76}},\ \bibinfo {pages} {035411} (\bibinfo {year}
  {2007})%
  \bibAnnoteFile{NoStop}{Tworzydlo:PRB07}%
\bibitem{Akhmerov:PRB08a}%
  \BibitemOpen
  \bibfield{author}{%
  \bibinfo {author} {\bibfnamefont{A.~R.}\ \bibnamefont{Akhmerov}}\ and\
  \bibinfo {author} {\bibfnamefont{C.~W.~J.}\ \bibnamefont{Beenakker}},\ }%
  \bibfield{journal}{%
  \bibinfo {journal} {Phys. Rev. B}\ }%
  \textbf{\bibinfo {volume} {77}},\ \bibinfo {pages} {085423} (\bibinfo {year}
  {2008})%
  \bibAnnoteFile{NoStop}{Akhmerov:PRB08a}%
\bibitem{Datta:97}%
  \BibitemOpen
  \bibfield{author}{%
  \bibinfo {author} {\bibfnamefont{S.}~\bibnamefont{Datta}},\ }%
  \emph{\bibinfo {title} {Electronic transport in mesoscopic systems}}\
  (\bibinfo {publisher} {Cambridge Univ Press},\ \bibinfo {year} {1997})%
  \bibAnnoteFile{NoStop}{Datta:97}%
\bibitem{Metalidis:PRB05}%
  \BibitemOpen
  \bibfield{author}{%
  \bibinfo {author} {\bibfnamefont{G.}~\bibnamefont{Metalidis}}\ and\ \bibinfo
  {author} {\bibfnamefont{P.}~\bibnamefont{Bruno}},\ }%
  \bibfield{journal}{%
  \bibinfo {journal} {Phys. Rev. B}\ }%
  \textbf{\bibinfo {volume} {72}},\ \bibinfo {pages} {235304} (\bibinfo {year}
  {2005})%
  \bibAnnoteFile{NoStop}{Metalidis:PRB05}%
\bibitem{Prada:PRB10}%
  \BibitemOpen
  \bibfield{author}{%
  \bibinfo {author} {\bibfnamefont{E.}~\bibnamefont{Prada}}, \bibinfo {author}
  {\bibfnamefont{P.}~\bibnamefont{San-Jose}}, \bibinfo {author}
  {\bibfnamefont{G.}~\bibnamefont{Le\'on}}, \bibinfo {author}
  {\bibfnamefont{M.~M.}\ \bibnamefont{Fogler}},\ and\ \bibinfo {author}
  {\bibfnamefont{F.}~\bibnamefont{Guinea}},\ }%
  \bibfield{journal}{%
  \bibinfo {journal} {Phys. Rev. B}\ }%
  \textbf{\bibinfo {volume} {81}},\ \bibinfo {pages} {161402} (\bibinfo {month}
  {Apr}\ \bibinfo {year} {2010})%
  \bibAnnoteFile{NoStop}{Prada:PRB10}%
\bibitem{Kim:EL08}%
  \BibitemOpen
  \bibfield{author}{%
  \bibinfo {author} {\bibfnamefont{E.-A.}\ \bibnamefont{Kim}}\ and\ \bibinfo
  {author} {\bibfnamefont{A.~H.~C.}\ \bibnamefont{Neto}},\ }%
  \bibfield{journal}{%
  \bibinfo {journal} {Europhys. Lett.}\ }%
  \textbf{\bibinfo {volume} {84}},\ \bibinfo {pages} {57007} (\bibinfo {year}
  {2008})%
  \bibAnnoteFile{NoStop}{Kim:EL08}%
\bibitem{Gomez-Medina:PRL01}%
  \BibitemOpen
  \bibfield{author}{%
  \bibinfo {author} {\bibfnamefont{R.}~\bibnamefont{G{\'o}mez-Medina}},
  \bibinfo {author} {\bibfnamefont{P.}~\bibnamefont{San-Jose}}, \bibinfo
  {author} {\bibfnamefont{A.}~\bibnamefont{{Garc{\'\i}a-Mart{\'\i}n}}},
  \bibinfo {author} {\bibfnamefont{M.}~\bibnamefont{Lester}}, \bibinfo {author}
  {\bibfnamefont{M.}~\bibnamefont{Nieto-Vesperinas}},\ and\ \bibinfo {author}
  {\bibfnamefont{J.~J.}\ \bibnamefont{S{\'a}enz}},\ }%
  \bibfield{journal}{%
  \bibinfo {journal} {Phys. Rev. Lett.}\ }%
  \textbf{\bibinfo {volume} {86}},\ \bibinfo {pages} {4275} (\bibinfo {year}
  {2001})%
  \bibAnnoteFile{NoStop}{Gomez-Medina:PRL01}%
\end{thebibliography}%

\setcounter{figure}{0}
\renewcommand\thefigure{S\arabic{figure}} 

%\widetext
\section{Supplementary material}

\begin{figure}[h!]
\includegraphics[width=8cm]{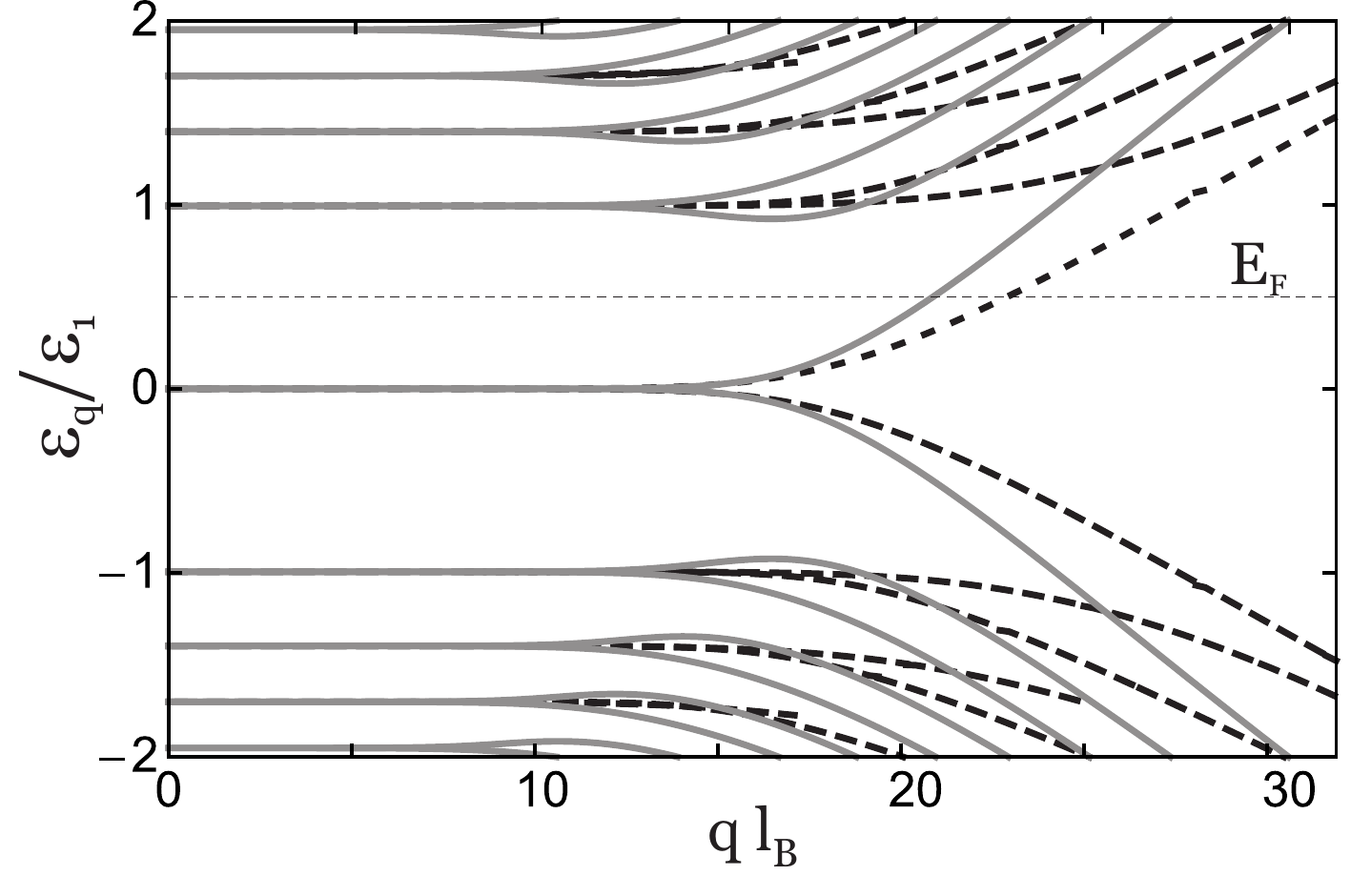}
\caption{The bands of an infinite nanoribbon with armchair (solid) and zigzag (dashed) boundary orientations. The Fermi energy (dotted) intersects the dispersive edge state branches, which fixes the edge state momentum q along the boundary.}\label{fig:bands}
\end{figure}

\subsection{Fig. \ref{fig:bands} : Nanoribbon bandstructure in the Hall regime}

According to Bloch's theorem, an infinite graphene nanoribbon generated by the repetition of a unit supercell of width $W$, has eigenstates that are plane waves of momentum $q$ in the direction $x$ along the nanoribbon. The dispersion relation in the presence of strong magnetic field (shown Fig. \ref{fig:bands} for forward propagating states) is composed of flat subbands for $q$'s that correspond to states localized away from the boundaries (most similar to bulk Landau levels), that become dispersive as the states move closer to the boundaries. When that happens, the valley degeneracy of the bulk Landau levels is lifted at the same time as the edge state branches become dispersive. The zigzag and armchair boundaries do so in slightly different ways, although both acquire a well defined (position independent) valley polarization. One essential feature, however, is independent of the boundary details: the Landau level at the Dirac point (zero energy) is split into opposite energy branches, so that the corresponding edge state brances intersect the Fermi energy only once. This is in contrast to all the other Landau levels, which contribute with twice the number of edge states. The resulting Hall conductivity $\sigma_{xy}=G_0(2N+1)$ (equal to the two terminal ballistic conductance $G$) in graphene monolayer nanoribbons is quantized at odd multiples of $G_0=2 e^2/h$, where $N$ is the number of populated subbands above the Dirac point. The resulting shift at $N=0$ is the hallmark of graphene's anomalous QHE.
\break

\begin{figure}[h!]
\includegraphics[width=7cm]{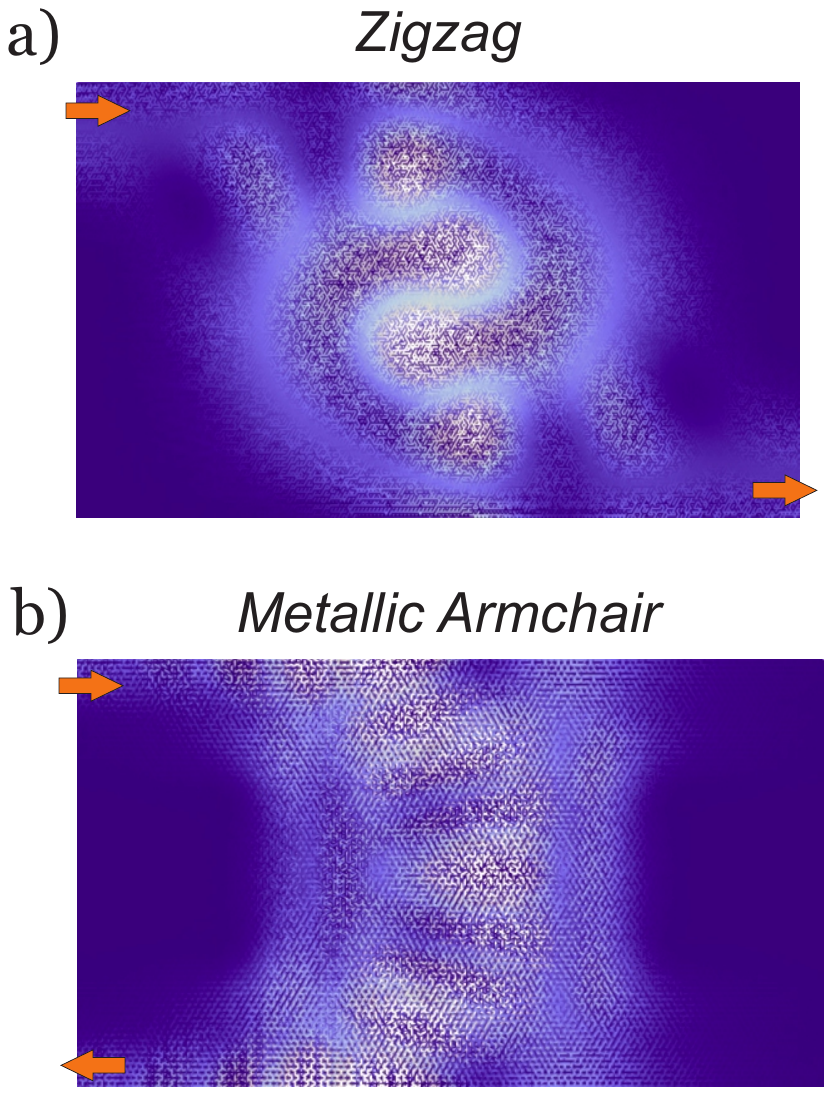}
\caption{(Color online) Charge density around the fold for incoming edge state energies slightly below the $N=1$ plateau, where evanescent modes from the latter dominate the electronic density. The parameters used for these simulations are: a)$W=81\textrm{nm}$, $R=4\textrm{nm}$, $l_B=12.2\textrm{nm}$ ($B=4.4\textrm{T}$), and energy $\epsilon=68\textrm{meV}$; b)$W=81\textrm{nm}$, $R=19\textrm{nm}$, $l_B=8.1\textrm{nm}$ ($B=10\textrm{T}$), and energy $\epsilon=96\textrm{meV}$. In b) the energy lies at a quasi-bound state in a metallic armchair nanoribbon, causing resonant backscattering. In the zigzag case above, resonant scattering does not induce backscattering due to the valley degeneracy of quasi-bound states.}\label{fig:quasibound}
\end{figure}

\subsection{Fig. \ref{fig:quasibound} : Quasibound states in folded nanoribbons}

\begin{figure}
\includegraphics[width=8cm]{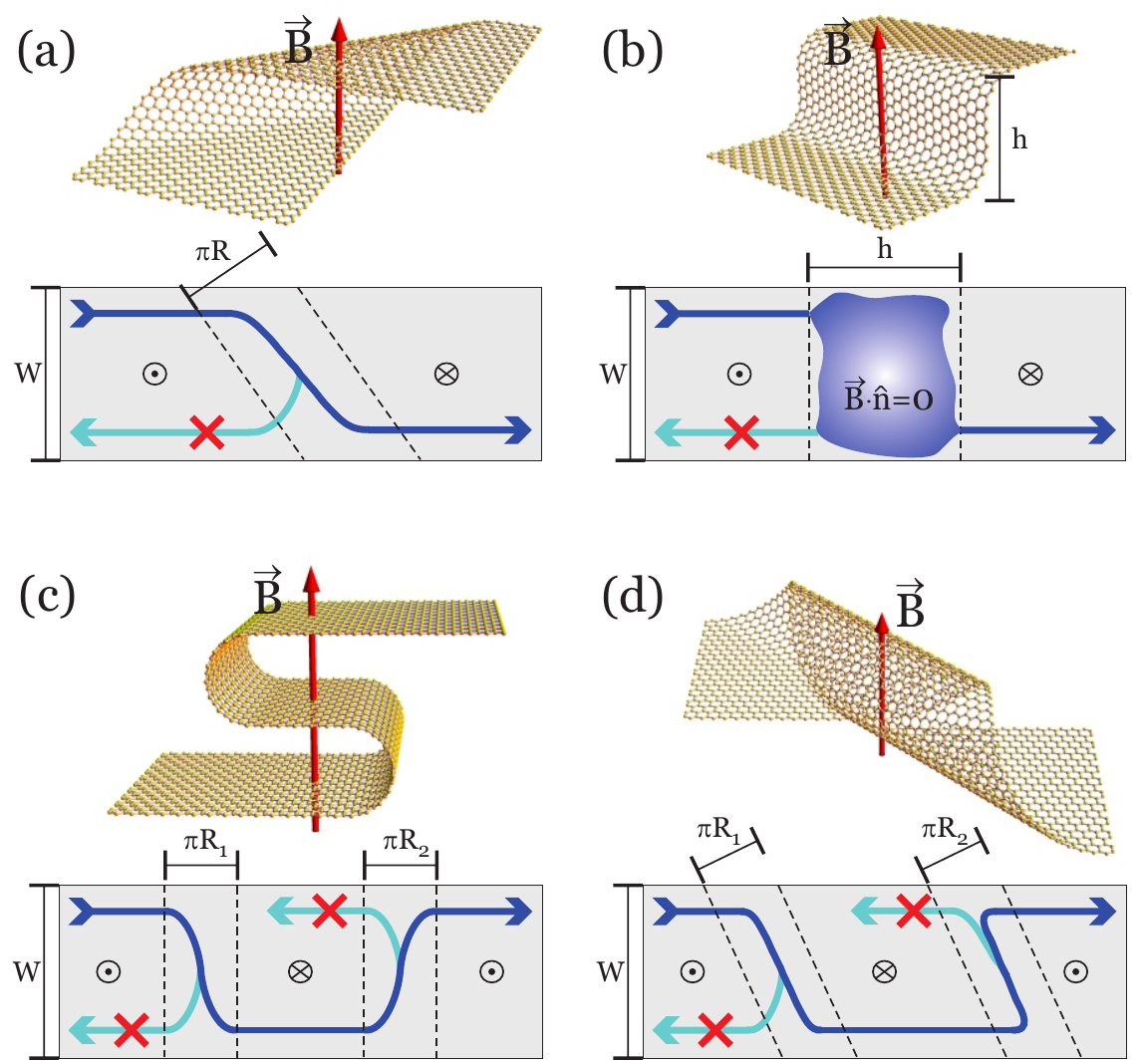}
\caption{(Color online) Variations of the folded nanoribbon geometry considered in the text that might be produced experimentally: non-orthogonal folds (a), steps (b), double folds (c) and loops (d). Underneath each geometry, sketch of the spatial charge density of the corresponding scattering states.} \label{fig:geometries}
\end{figure}

\begin{figure}
\includegraphics[width=8cm]{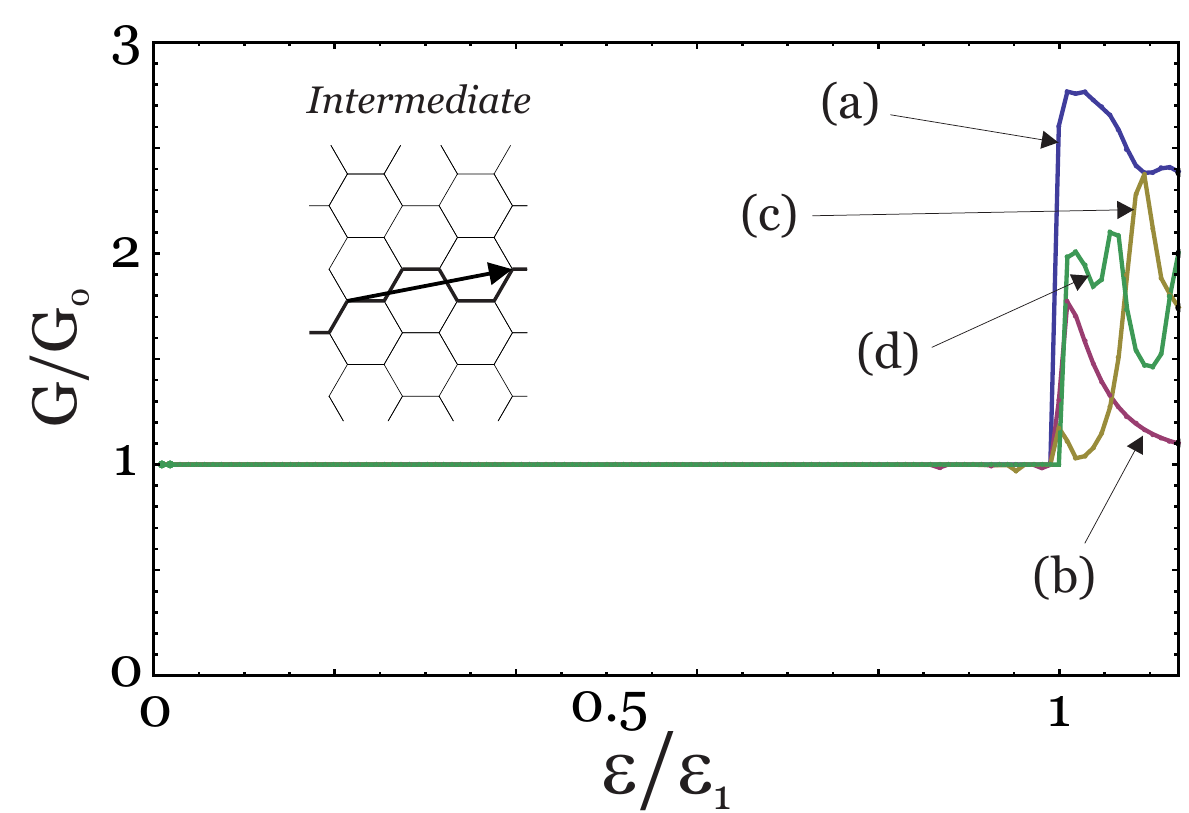}
\caption{(Color online) Computed conductance across the four fold variations of Fig. \ref{fig:geometries}  for system parameters like in Fig. 3 of the main text. On the first Landau plateau, the transmission is perfect, like in the orthogonal fold geometry. Crystallographic orientation is intermediate $\theta=0.1\pi/6$, but similar results were obtained for zigzag and armchair terminations.} \label{fig:Tgeometries}
\end{figure}

\subsection{Figs \ref{fig:geometries} and \ref{fig:Tgeometries} : Non-orthogonal folds, and other geometry variations}

Experimentally, graphene nanoribbon folds may be produced that deviate from the fold geometry studied in the main text, see Fig. \ref{fig:geometries}. In particular, folds that are non-orthogonal to the ribbon axis are possibly the easiest to produce by water jet folding techniques, and are better candidates to implement transport experiments, since it is easier to contact the two layers to different reservoirs. Since the precise fold geometry only affects the scattering wavefunction around the fold, and not the asymptotic edge states, the transmission is perfect, like in the orthogonal fold case, see Fig. \ref{fig:Tgeometries}.

\begin{figure}
\includegraphics[width=8cm]{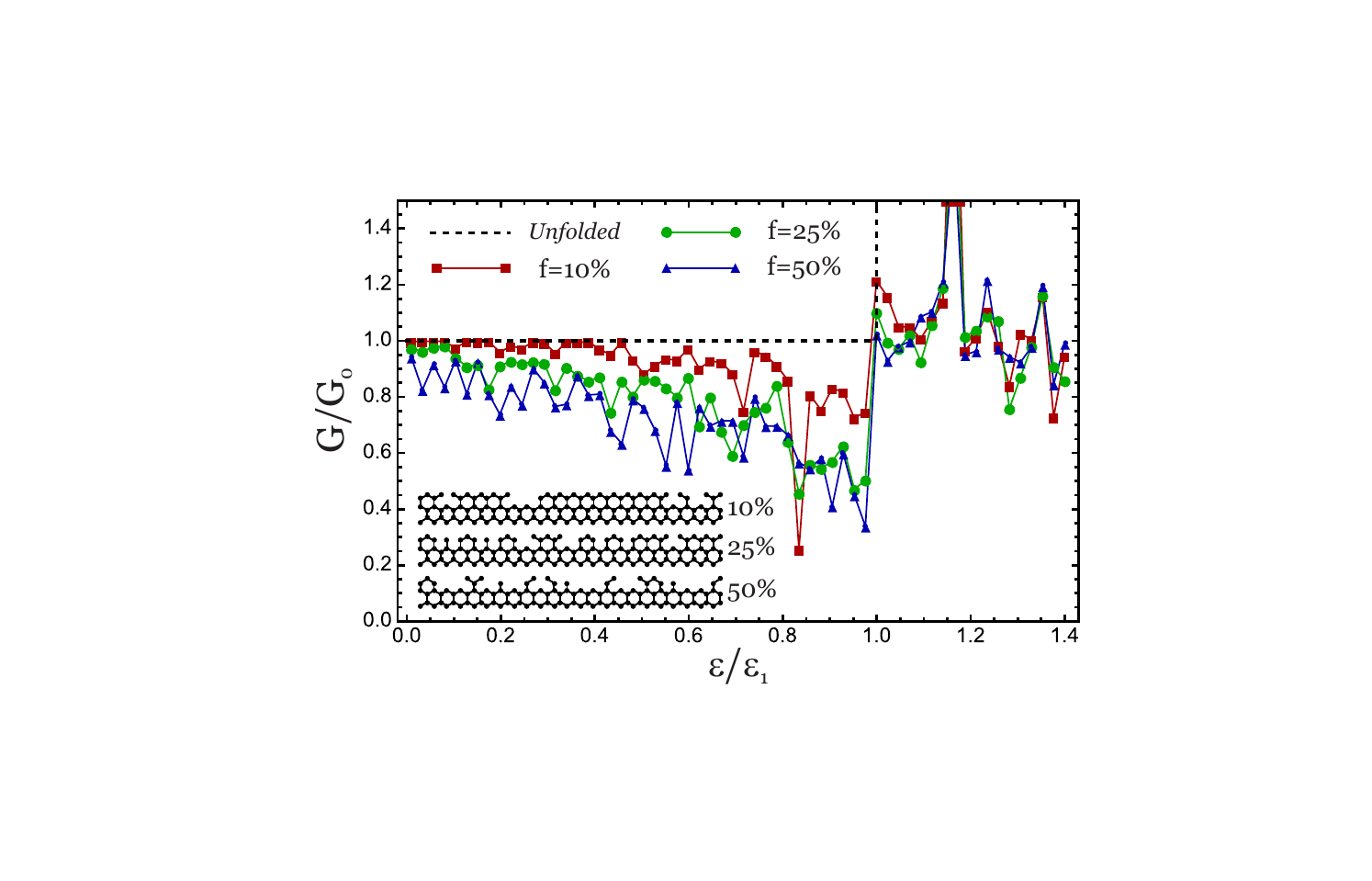}
\caption{(Color online) Two-terminal conductance (normalized to the conductance quantum and averaged over $16$ samples) as a function of the edge state energy (normalized to the  first Landau level energy) in the presence of shallow edge roughness. The clean nanoribbon is zigzag terminated and has parameters: $W = 81\textrm{nm}$, $R = 20.3\textrm{nm}$ and $l_B = 8.1\textrm{nm}$ ($B = 10\textrm{T}$). Boundary disorder is parametrized in terms of the fraction of vacancies $f$ along the last two atomic layers of the boundary (as shown schematically in the inset, which shows the nanoribbons unfolded flat for the purpose of these plots). $f=10\%$ for the squared (red) curve, $f=25\%$ for the circled (green) one and $f=50\%$ for the triangled (blue) one. The case of an unfolded clean zigzag nanoribbon (black dashed curve) is shown for comparison.}\label{fig:disorder}
\end{figure}

\subsection{Fig. \ref{fig:disorder} : Disorder in the boundary termination}
Transport in the presence of edge roughness is numerically computed for the folded nanoribbon of Fig. 3 with zigzag termination. A fraction $f$ of individual atomic sites are removed from the last two boundary layers, modeling shallow edge roughness.  In Fig. \ref{fig:disorder} the resulting average interlayer transmission is plotted for increasing fraction $f$. It is apparent that, especially at vanishing energy for which the edge state moves a certain distance into the bulk, the edge roughness (assumed smaller than $l_B$, the typical edge state width) has a small effect on the transmission.

\break

\end{document}